# Thermal Fluctuation-Induced Electricity Generation across a Non-Ideal Diode

Guoan Tai,[1,2] Jinsong Liu,[1,3] Tian Zeng,[1,3] Jizhou Kong,[1,4] and Fuyong Lv[5]

[1]The State Key Laboratory of Mechanics and Control of Mechanical Structures, Nanjing University of Aeronautics and Astronautics, Nanjing 210016, China

[2]College of Aerospace Engineering, Nanjing University of Aeronautics and Astronautics, Nanjing 210016, China

[3]School of Material Science and Technology, Nanjing University of Aeronautics and Astronautics, Nanjing 210016, China

[4]College of Mechanical and Electrical Engineering, Nanjing University of Aeronautics and Astronautics, Nanjing 210016, China

[5]School of Electronic Science and Engineering, Southeast University, Nanjing 210096, China

**We report experimentally that the electricity is generated from ambient thermal fluctuations across a non-ideal diode consisted of a silicon tip and an aluminum surface. The output is tuned by the contact force which modulates Schottky barrier heights as well as rectifying ratios of the diodes. The interaction regime between the silicon and the aluminum locates at the quantum-classical boundary where thermal fluctuations are appreciable, and the rectification of thermal fluctuations leads to the electricity generation. This finding offers an innovative approach to environmental energy harvesting.**

PACS numbers: 05.70.Ln, 05.40.Ca, 07.50.Hp, 05.70.-a

[*] To whom correspondence should be addressed. E-mail: taiguoan@nuaa.edu.cn

As random deviations of a system from its average state, rectifying thermal fluctuations at the macroscale are believed to be impossible under thermal equilibrium conditions, nevertheless the case is completely different at a small scale [1-4]. As a system's spaces or dimensions decrease, the thermal fluctuations result in the system deviating from the thermodynamic equilibrium, and some useful and unexpected properties are usually predicted or observed in a nonequilibrium system [1,3]. For instance, extracting useful work from a single heat bath was predicted theoretically in quantum regime [5-7], and the erasure of quantum information has been ascribed experimentally to a source of entropy loss in the nonequilibrium system [8-10].

The decrease of entropy in the above systems were always explained by the rectification of microscopic thermal fluctuations as mentioned in many variants of the Maxwell's demon, such as thermal ratchets, Szilard's engine, quantum engine and the feedback control of microscopic fluctuations [5-16]. The nonequilibrium thermal fluctuations are appreciable at a small scale so that manipulating such small devices is very important for rectifying and harvesting the thermal fluctuations. Up to now, direct energy harvesting from thermal fluctuations in a small-scale device is not still reported experimentally in the absence of the external energy input.

In this work, we report a process of harvesting energy from the ambient thermal fluctuations and transforming them into electricity by metal-semiconductor contacts. The output strikingly depends upon the contact force between the silicon tip and the aluminum surface. Weak asymmetry in the non-ideal diodes induces a nonequilibrium thermodynamic environment, leading to the electricity generation. This finding offers an innovative approach to environmental energy harvesting.

Schematic diagram of the device and corresponding circuit were shown in Fig. 1(a), where the aluminum (Al) ingot was taken as the planar metal, and the lightly doped n-silicon (Si) (marked as n-Si(4)) was used as the semiconductor tip [17]. The phase structure of the Al and the Si was examined by XRD analyses [17]. The output voltage ($V_{op}$) of the device with a 100 kΩ load and an applied force of 0.61 N fluctuated from 0.8 to 2.8 mV over 34 h under different atmospheric conditions including $N_2$ gas, vacuum and air in a black and sealed metal box (Fig. 1(b)), and the results show that the output is independent of the atmospheric conditions. The corresponding open-circuit voltage ($V_{oc}$) in vacuum fluctuated from 1.8 to 4.8 mV for over 10 h (Fig. 1(c)).

The Al-Si contact degree was quantitatively characterized by the contact force. The $V_{op}$ values under 0.03-9.16 N contact force was shown in Fig. 2(a), the higher $V_{op}$ (0.53-2.50 mV) was observed at the contact force between 0.05 and 2.22 N, and the corresponding $V_{oc}$ displayed the same trend (Fig. 2(b)). The Al-Si contact characteristics were further evaluated by current-voltage (*I-V*) curves, as shown in Fig. 2(c). The asymmetry in the *I-V* curves is gradually removed with increasing contact force from 0.03 to 9.16 N, implying the decreasing of the Schottky barrier heights (SBHs), which is demonstrated by the calculated SBHs based on the thermionic emission theory [17]. The SBHs were closely correlated with the effective contact area ($S_{eff}$) [18,19], and the $S_{eff}$ was calculated by the Hertz theory [20] and then the penetration depth ($D_p$) was obtained, as shown in Fig. 2(d). In combination with Fig.

2(a), the output voltage was markedly higher than 0.53 mV when the $D_p$ is less than 1 μm which locates at the quantum-classical boundary, thus quantum decoherence is responsible for the electricity generation [5,6,21,22]. Subsequently, the SBHs and rectifying ratios were found to modulate from 0.624 to 0.054 eV and from 2.28 to 1.68, respectively (Fig. 2(e)) [17]. Compared with the rectifying ratio of an intimate standard diode (beyond $10^7$) and an ideal Ohmic contact ($10^0$) [18], the experimental results suggest that we obtained the non-ideal diodes with tuning SBHs and rectifying ratios. The non-ideal diodes can be expressed by an equivalent circuit with the series and parallel resistances, and the resistances decrease with increasing the contact force (Fig. 2(f)) [17].

To ascertain why the continuous output was spontaneous in the non-ideal diodes, several experiments were performed. (1) The output of the Al-Si device was the same when the light was on and off [17]. (2) The output of a 100 kΩ resistor was close to zero, which excluded the possibility of the electricity generation from the thermal noise of the resistor [17]. (3) Temperature-dependent output data showed that the output was almost no difference under various temperatures, which results from a slight barrier height change with temperature [17-19]. Therefore, the spontaneously continuous output could be ascribed to an ambient thermal fluctuation in the non-ideal diodes.

To reveal the underlying physical mechanism for the electricity generation, an ambient thermal fluctuation induced self-charging mechanism is therefore proposed to elucidate our experimental results. It is well-known that the free electrons with concentrations of up to $10^{10}$/cm$^3$ on an intrinsic semiconductor Si surface are usually under thermal fluctuations with high speed (the most probable speed, $v_p = (2k_BT$

/m$^*$)$^{1/2}$≈10$^5$ m/s) at room temperature [18,19]. Probability distribution of the thermal fluctuations suggests that there are always some thermal electrons with high energy to go through the energy barrier, and the amount of the thermal electrons through the barrier strongly depends upon the barrier height [18,19].

For a non-ideal diode, previous theories implied that the temperature of the resistor is different from that of the diode, but the temperature difference is negligible because of far longer time scale of temperature relaxation ($\tau_T$) than that of the heating time ($\tau_H$) [23]. Experimentally, the energy harvesting process occurs spontaneously without external bias voltage and temperature difference, and the continuous output implies that the process involves a transition from classical equilibrium thermodynamics to nonequilibrium thermodynamics.

For a classically intimate standard diode, the strict thermodynamic equilibrium state is well-established because of exactly balancing the electrostatic force by the diffusive force, where a depletion layer is invariably formed between a semiconductor and a metal or another semiconductor surfaces [18,19]. The layer blocks the flowing of the thermal electrons and builds up the barrier for modulating the electric transport by an applied electric field, and the barrier's energy difference is embodied as the SBH [18,19]. The depletion width ($W_D$), namely the width of the depletion region in a semiconductor, which is governed by the principle of charge neutrality, can be expressed as [18]

$$W_D = \sqrt{\frac{2\varepsilon_{Si}\varepsilon_0}{qN_d}(\psi_{bi} - V_a - \frac{k_B T}{q})} \qquad (1)$$

where $\varepsilon_{Si}$ is the relative dielectric constant of the Si, $\varepsilon_0$ is the dielectric constant of vacuum, $q$ is the elementary charge, $N_d$ is the impurity concentration, $\psi_{bi}$ is the built-in potential, $V_a$ is the applied electric field, $k_B$ is Boltzmann constant and $T$ is the absolute temperature. The $\psi_{bi}$ is $\phi_{B0} - (k_B T)\ln(N_c / N_d)$, where $N_c$ is the effective density of states in the conduction band of the n-Si ($N_c = 2.81 \times 10^{19}$ cm$^3$). When $V_a$ is zero, we calculated a series of $W_D$ based on Eq. (1) under different $\psi_{bi}$ values [17].

For the non-ideal Al-Si diode circuit (Fig. 3(a)), thermal fluctuation (TF) force in the system breaks the balance between the electrostatic force and the diffuse force, leading to the establishment of a nonequilibrium thermodynamic environment (Fig. 3(b)). Thus, the trend toward the thermodynamic equilibrium state and the persistently thermal fluctuation induce the energy harvesting. The corresponding equivalent circuit of the self-charging and self-discharging circle is proposed, where the self-charging process is maintained by the thermal fluctuations of electrons at the Si tip and the self-discharging behavior arises from the recovery toward the thermodynamic equilibrium state through the resistances (Fig. 3(c)), which is similar to that of our recent finding for the electricity generation from ionic thermal motion across a silicon surface [24].

To understand in detail the transition between the equilibrium and the nonequilibrium thermodynamics during the Al-Si contact, a two-dimensional Atlas electronic device numerical simulation was performed to obtain the potential distributions around the junctions, the *I-V* curves and the energy level diagrams. Three typical states were modeled: (i) a high SBH up to 0.81 eV when $D_p$ is zero (Fig. 3(d)); (ii) a moderate SBH equal to 0.38 eV when $D_p$ is 100 nm (Fig. 3(e)); (iii) a low SBH down to 0.11 eV when $D_p$ is 3 μm (Fig. 3(f)). Equipotential distributions near to the junctions are shown in Fig. 3(d-f) (top trace), and the dense equipotential lines, especially at the edge of the contact, indicate a high electric field [17]. The *I-V* curve exhibits that state (i) is similar to a standard planar diode with a strong asymmetry, where the device is in a strict thermodynamic equilibrium (down trace, Fig. 3(d)) [17]. When the $D_p$ is less than one micrometer, the *I-V* curve of a non-ideal diode characteristic with the rectifying ratio down to 8 shows a weak asymmetry (down trace, Fig. 3(e)). Furthermore, after the $D_p$ is increased to three micrometers, the *I-V* curve with a symmetry suggests that the Ohmic contact is formed at the junction (Fig. 3(f)). The simulated results are in good agreement with our experiments.

From the above, we know that the thermal fluctuations at the junction cannot be neglected for the nonequilibrium state located at the sub-micron interaction region. The junction situated in a thermal bath can be regarded as a small capacitor so that the thermal fluctuations of electrons inevitably change its voltage difference [25-28]. The averaged square of the thermal fluctuation voltage across the junction, $\langle V_{TF}^2 \rangle$, can be calculated as [26,28]

$$\langle V_{TF}^2 \rangle = \int_0^\infty \frac{4k_B TR \mathrm{d}f}{(2\pi fCR)^2 + (1+C/C_0)^2} \tag{2}$$

where *R* is the resistance of the equivalent resistor, *f* and d*f* are the frequency and the frequency interval of the resistor, respectively, and *C* and $C_0$ are the equivalent capacitance of the diode and the total capacitance of the diode, respectively.

The $C/C_0$ term can be close to zero because *C* is far less than $C_0$, so we have

$$\langle V_{TF}^2 \rangle = k_B T / C \tag{3}$$

The capacitance *C* is given by

$$C = \varepsilon_{Si}\varepsilon_0 \frac{S_{eff}}{d_{eff}} = \varepsilon_{Si}\varepsilon_0 \frac{S_{eff}}{W_{TF}} = \varepsilon_{Si}\varepsilon_0 \frac{S_{eff}}{\xi W_D} \tag{4}$$

where $S_{eff}$ is the effective area of the junction, and $d_{eff}$ is the barrier width (equal to the depletion width of the junction in a standard Schottky diode), $W_{TF}$ is the width of a dynamical energy drop region in a nonequilibrium thermal fluctuation, $W_D$ is the depletion width in a strict thermodynamic equilibrium, and $\xi$ is the thermal fluctuation factor ($\xi = W_{TF}/W_D$). The $\xi$ is unity for a strict thermodynamic equilibrium system, whereas it is usually larger than unity in a nonequilibrium thermodynamic one.

The $C$ reduces with decreasing $S_{eff}$ based on Eq. (4), and the small $C$ makes the thermal voltage fluctuations appreciable when the junction is at the sub-micron scale [25-28]. The fluctuating thermal voltage (Unit: mV) was obtained by

$$V_{TF} = \left(\frac{k_B T W_{TF}}{S_{eff}\varepsilon_{Si}\varepsilon_0}\right)^{1/2} = 0.00632\sqrt{\xi W_D / S_{eff}} \qquad (5)$$

Here, the $V_{TF}$ is equal to the measured $V_{oc}$ because the electricity generation arises from the thermal fluctuations. So, we calculated a series of the $\xi$ versus the contact force according to Eq. (5) [17]. To investigate the feasible range of the $\xi$ with respect to the high output, we define the pressure-related geometric factor $K_{PF} = 10^{-6} F/(S_{eff} S_{total})$. The $\xi$ is well-fitted as a function of $K_{PF}$ by a Lorentz function $\xi = -4.591 + 4.835/(4(K_{PF}-0.357)^2 + 0.056)$ (Fig. 4(a)), and the higher $\xi$ dispersing around $K_{PF}=0.357$ benefits the higher output, as the $V_{op}$-$K_{PF}$ curve shown in Fig. 4(b). Based on Eq. (5) and the equivalent circuit in Fig. 3(c), the output voltage ($V_{op,theory}$) is given by

$$V_{op,theory} = \frac{0.00632 R_{ex}}{R_{in} + R_{ex}}\sqrt{\xi W_D / S_{eff}} \qquad (6)$$

where $R_{in}$ is the internal resistance of the non-ideal diode, $1/R_{in} = 1/R_S + 1/R_L$.

Comparing the measured $V_{op,exp}$ with the calculated $V_{op,theory}$, we found that the proposed theory is in excellent agreement with the experimental results (Fig. 4(b))

[17]. The power density of the devices can be estimated by $W = V_{oc}^2 / (R_{in} \cdot S_{eff})$, where $W$ is the electrical power. The calculated maximum power density from our device is 32.3 μW/cm$^2$ under an applied force of 0.51 N. In addition, further studies showed that the output of the Al-Si devices strongly depended upon the doping concentration of the Si wafers [17]. Moreover, when the Al was substituted by other metals, the output was independent on their work functions, which is due to the pinning of the Fermi levels at the surfaces [17-19].

In conclusion, we have demonstrated the electricity generation from thermal fluctuation of electrons across a non-ideal metal-semiconductor diode under room temperature. The output is effectively tuned by the contact force which modulates the Schottky barrier heights as well as the rectifying ratios of the diodes. Quantum decoherence can be used to explain the electricity generation because the interaction regime locates at the quantum-classical boundary where thermal fluctuations are appreciable. The study builds a bridge between the classical equilibrium thermodynamics and the nonequilibrium thermodynamics, and offers an innovative approach to environmental energy harvesting.

We thank Y. Xiao, T. Xu, Z. Xu and T. Sun and for many fruitful discussions. This work was supported by NSF (11302100), Innovation Fund of NUAA (NS2013095 and NS2013060), the Research Funds of KL-INMD (NJ20140002), SKL-MCMS (0413Y02) and the Priority Academic Program Development of Jiangsu Higher Education Institutions.

**Figure legends**

FIG. 1 (color online). Experimental setup and output of the non-ideal Al-Si diode at room temperature. (a) Schematic diagram showing the experimental setup and the corresponding circuit. (b) Output voltage of the diode in $N_2$ gas, vacuum and air when a 100 kΩ resistor is in parallel loaded to the circuit. (c) Open-circuit voltage of the diode in vacuum. The applied force is 0.61 N.

FIG. 2 (color online). Contact force-dependent output characteristics of the non-ideal Al-Si diodes. (a) The contact force-dependent output voltage of the diodes when a 100 kΩ resistor is loaded to the circuit. (b) The contact force-dependent open-circuit voltage of the diodes. (c) The contact force-dependent *I-V* characteristics of the diodes. The inset is an enlarge view of the curves with the voltage range from 0 to 0.4 V. (d) The contact force-dependent effective contact areas ($S_{eff}$) as well as penetration depths ($D_p$). (e) The contact force-dependent Schottky barrier heights (SBHs) as well as rectifying ratios. (f) The contact force-dependent series resistance ($R_s$) as well as parallel resistance ($R_L$).

FIG. 3 (color online). Electricity generation mechanism of the non-ideal diodes. (a) A circuit diagram consisted of the Al-Si diode and the external load $R_{ex}$. (b) The detail of the Al-Si junction, where the thermal fluctuation force is proposed to elucidate the electricity generation. (c) Equivalent circuit indicating the charging and discharging circle. (d) Top view: Calculated potential distribution of the ideal Al-Si Schottky contact. Bottom view: *I-V* characteristic of the device showing a high SBH of up to 0.81 eV and a high rectifying ratio on the order of $10^7$ (**strong asymmetry**). (e) Top view: Calculated potential distribution of the non-ideal Al-Si Schottky contact with the $D_p$ of 100 nm. Bottom view: *I-V* characteristic of the device showing a moderate SBH of 0.38 eV and a weak rectifying ratio of approx. 8 (**weak asymmetry**). (f) Top view: Calculated potential distribution of the ideal Al-Si Ohmic contact with the $D_p$ of 3 μm. Bottom view: *I-V* characteristic of the device showing a low SBH of 0.11 eV and a rectifying ratio of unity (**strong symmetry**). The insets in (d)-(f) are the corresponding energy band diagrams.

FIG. 4 (color online). Pressure-related geometric factor dependence of thermal fluctuation parameters. (a) The thermal fluctuation factor ($\xi$) as a function of the pressure-related geometric factors ($K_{PF}$). The red line is a Lorentz fitting curve, and the horizontal dotted line is the thermodynamic equilibrium baseline. (b) Schottky barrier heights (SBHs) and rectifying ratios as functions of the pressure-related geometric factors. The SBHs and rectifying ratios of the ideal Schottky diode are set to be 0.81 eV and $10^7$, respectively, whereas those of the ideal Ohmic contact are 0 eV and zero, respectively.

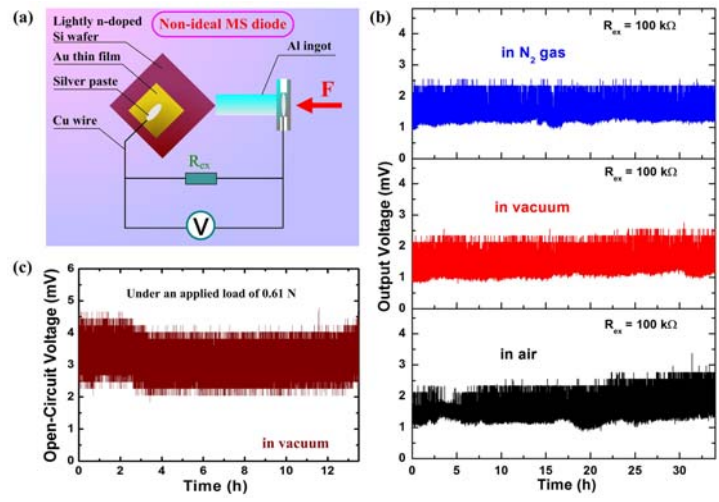

FIG.1

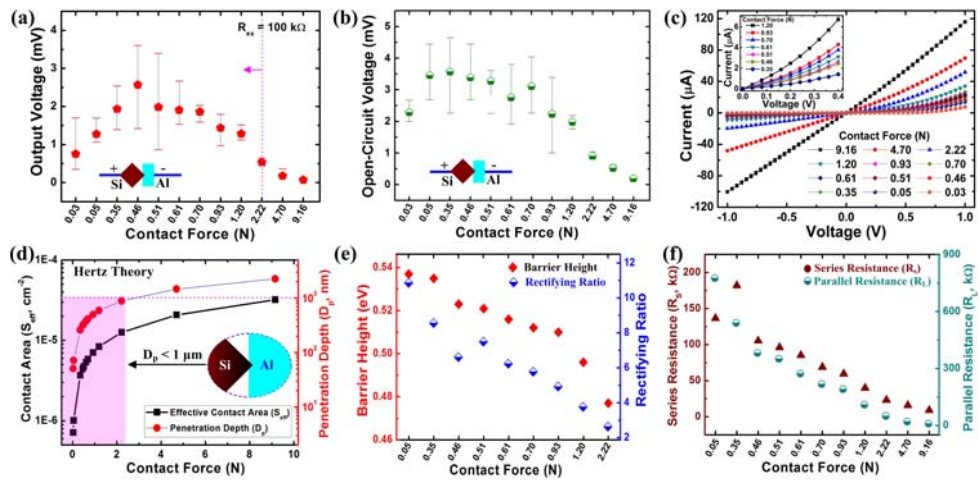

FIG. 2

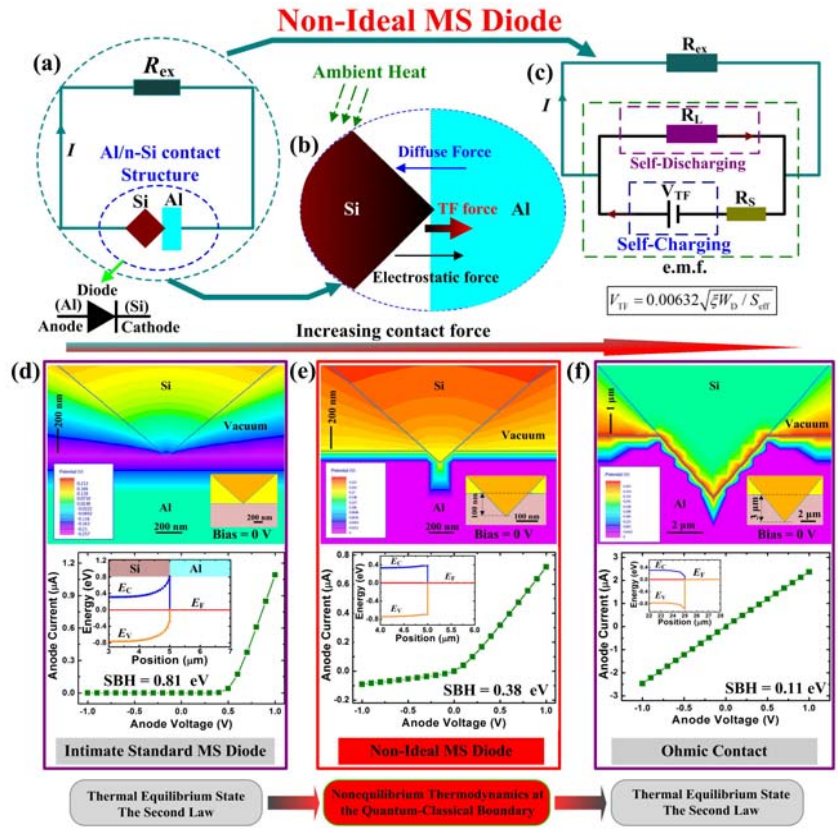

FIG. 3

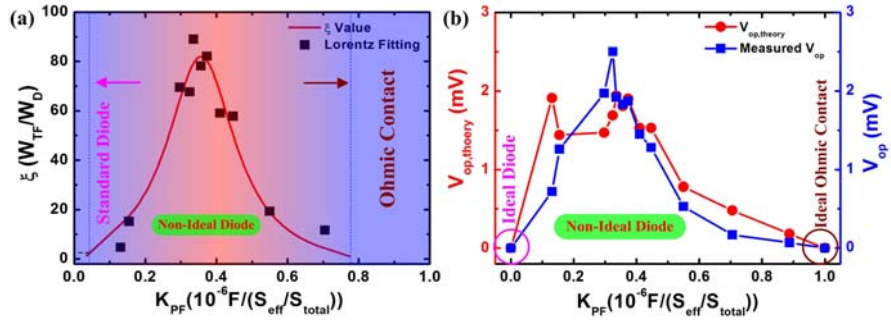

FIG. 4